\documentclass[useAMS,natbib]{mn2e}
\usepackage{amssymb,amsmath}
\usepackage{graphicx}

\title[Spectral index - luminosity relation]
{Optical spectral index - luminosity relation for the 17 mapped 
  Palomar-Green quasars}

\author[Zhang]
      {Xue-Guang Zhang$^{1,2}$\thanks{xgzhang@pmo.ac.cn}\\
       $^1$Purple Mountain Observatory, Chinese Academy of Sciences,
             2 Beijing XiLu, NanJing, JiangSu, 210008, People's Republic 
             of China \\
       $^2$Chinese Center for Antarctic Astronomy, NanJing,
             JiangSu, 210008, People's Republic of China}

\date{}

\pagerange{\pageref{firstpage}--\pageref{lastpage}} \pubyear{}
\def\LaTeX{L\kern-.36em\raise.3ex\hbox{a}\kern-.15em
   T\kern-.1667em\lower.7ex\hbox{E}\kern-.125emX}
\begin{document}
\label{firstpage}

\maketitle
\begin{abstract}
   In this paper, the optical spectra index - luminosity relationship 
is checked for the well-known 17 individual mapped QSOs, in order to give 
one more clearer conclusion on the so far conflicting dependence of the 
spectral index on the luminosity for AGN. Different from the global 
relationships based on the color difference (photometry parameters) for 
samples of AGN, the more reliable relationship is determined for the 
multi-epoch observed individual mapped QSOs with no contamination from the 
host galaxies, the line variabilities and the much different central 
properties. The final confirmed results are as follows. (1): No strong 
dependence of the optical spectral index on the continuum luminosity can 
be found for all the 17 QSOs, besides two objects (PG 0026 and PG 1613) 
having some weak trends (with $3\sigma$ confidence level) for the 
relationship. In other words, the common sense 'AGNs get bluer when they 
get brighter' is not so common. (2): There are much different damped 
intrinsic variability time scales for the variability modes of the optical 
spectral index and the continuum emission, through the well applied 
Damped Random Walk method for the AGN variability. In other words, there 
are some different intrinsic mechanisms controlling the variabilities of 
the optical spectral index and the power law AGN continuum emission. 
Therefore, the much weak dependence of the optical spectral index on 
the continuum luminosity can be further confirmed. 
\end{abstract}


\section{Introduction}
    Variability is one of the fundamental characters of Active Galactic 
Nuclei (AGN) (Ulrich et al. 1997), and moreover, AGN variability is one 
powerful tool for understanding the structures of the central regions of 
AGN (Rees 1984, Hawkins 1996, Kawaguchi et al. 1998, 
Torricelli-Ciamponi et al. 2000, Hawkins 2002, Peterson et al. 2004, 
Hopkins et al. 2006, Kelly et al. 2009, Breedt et al. 2010, 
Schmidt et al. 2010, Mushotzky et al. 2011, Zhang 2011, Zu et al. 2011, 
Zhang 2013a, 2013b, Zu et al. 2013). However, there is so far no clear 
conclusion about the nature of the AGN variability, besides several 
well-known dependence of the AGN variability on the other AGN parameters 
(such as the luminosity, accretion rate, black hole mass, redshift etc.., 
Vanden Berk et al. 2004, Wold et al. 2007, Wilhite et al. 2008). Among the 
dependence, the dependence of the spectral index (in the paper, we 
mainly consider the continuum slope 
$\alpha$, $f_{\lambda}\propto\lambda^{\alpha}$) on the AGN luminosity (the 
spectral index - luminosity relation) is the widely and commonly studied 
one, since the dependence was found for 3CR extragalactic galaxies 
(Heeschen 1960, Conway et al. 1963, Kellermann et al. 1969, 
Macleod \& Doherty 1972). However, the conclusion about the dependence 
is still uncertain.

    In the previous studies, the spectral index - luminosity relationship 
was commonly checked by samples of AGNs with different redshifts, different 
luminosities, different black hole masses, different accretion rates, 
different contamination sources etc.. Therefore, there are so far 
contradictory statements about the relationship.  More recently, 
Schmidt et al. (2012) have reported that AGN color variability (spectral 
index) is remarkably uniform, and independent not only of redshift, but also 
of quasar luminosity and black hole mass, after the correction of the 
effects from the emission lines in the different SDSS (Sloan Digital Sky 
Survey) bands, by the sample of 9093 QSOs in the SDSS stripe82. 
Zuo et al. (2012) have reported that AGN variability increases as 
either luminosity or Eddington ratio decreases, 
however, the relationship between variability and black hole mass is 
uncertain, by 7658 QSOs from SDSS Stripe 82.  Certainly, some different 
results about the spectral index - luminosity relationship can be still 
found in the literature. Sakata et al. (2010) have shown that the spectral 
shape of AGN continuum emission in the optical region does not systematically 
change during flux variation by the study of 11 nearby Seyfert galaxies. 
Woo et al. (2007) have shown that after the corrections of the host galaxy 
light dilutions, the nuclear variability with the similar amplitudes in the 
g and r bands within the errors, by the study of 13 moderate luminosity 
AGNs at $z=0.36$. The similar results can be found in Walsh et al. (2009): 
there are no significant color variations for the 13 nearby AGNs. Meanwhile, 
we (Zhang et al. 2008) have shown that there is one strong correlation 
between the spectral index and the dimensionless Eddington accretion 
rate by 193 broad line AGN with high quality SDSS spectra. Pu et al. 
(2006) have reported that there was strong global correlation between the 
spectral index and the continuum luminosity for the QSOs in Kaspi et 
al. (2000). Wilhite et al. (2005) have reported the clear dependence of 
the spectral variability on the wavelength by one sample of about 300 
variable QSOs with multi-epoch SDSS spectra. Therefore, there is so far  
no one confirmed conclusion about the spectral index - luminosity relationship.

    It is clear that the previous study of the color variability (
spectral index) was mainly based on photometry parameters of samples 
of AGN. Thus, there are apparent effects on the spectral index - 
luminosity relation from the contributions of the  host galaxies and 
the emission lines, and from the different AGN properties (different BH masses, 
redshift, accretion rate, etc) of the sample objects. Even though, 
some effects above can be statistically corrected as discussed in the 
corresponding literatures, the sample size should have apparent effects 
on the final conclusion about the spectral index -luminosity 
relationship (Tang et al. 2007). Therefore it is necessary and 
interesting to re-check the spectral index - luminosity relationship 
by the spectroscopic data of the well-known mapped objects. The 
advantages of using the spectroscopic parameters of the mapped objects 
are as follows. On the one hand, the effects of the different black hole 
masses can be totally ignored. On the other hand, the effects of the 
emission lines can be clearly ignored. Moreover, in the paper, 
we mainly consider the 17 well known mapped Palomar-Green QSOs (PG QSOs), 
thus, the effects of the host galaxies can be totally ignored.

   The paper is organized as follows. In section 2, we present our method 
to determine the spectral index of the mapped PG QSOs. Section 3 gives the 
spectral index - luminosity relationships for the mapped objects. Finally 
the discussions and conclusions are given in Section 4. In this paper, the 
cosmological parameters $H_{0}=70{\rm km\cdot s}^{-1}{\rm Mpc}^{-1}$, 
$\Omega_{\Lambda}=0.7$ and $\Omega_{m}=0.3$ have been adopted.

\section{The Spectral Indices of the 17 PG QSOs}

    In order to totally ignore the effects of the host galaxies and/or 
the probable effects of the flux calibration, the 17 nearby PG QSOs 
with public spectra (Kaspi et al. 2000) are mainly considered, among all 
the reported mapped AGN (kaspi et al. 1996, 2000, Peterson et al. 2004, 
Kaspi et al. 2005, Bentz et al. 2006, Denney et al. 2006, Bentz et al. 2009, 
Denney et al. 2009, Bentz et al. 2010, Denney et al. 2010, Barth et al. 2011). 
The detailed description of the 17 PG QSOs can be found in Kaspi et al. (2000). 
Their spectra were observed using the Steward Observatory 2.3 m telescope 
and the Wise Observatory (WO) 1 m telescope for around 7.5 years. Each 
QSO has more than 20 spectroscopic data points, which lead us to find 
the more reliable spectra index - luminosity relationship for the 17 
individual objects without the global effects. The public spectra have 
been collected from the website: http://wise-obs.tau.ac.il/\~shai/PG/. 
Moreover, the spectra have been binned into 1\AA per pixel and are 
padded from 3000\AA to 9000\AA. Certainly, in most cases the usable 
part of the spectrum is the blue part around the H$\beta$ 
(Maoz et al. 1994, Kaspi et al. 2000). Therefore, in this paper, the 
spectra around the H$\beta$ are mainly considered.

    Then, based on the multi-epoch observed spectra of the 17 mapped PG 
QSOs, the spectral index in the optical band (from 4400\AA to 5500\AA 
in the rest frame including the optical Fe~{\sc ii} lines) can be well 
determined by the following power law function, 
$f_{\lambda}\propto\lambda^{\alpha}$. Moreover, due to the effects of the 
optical Fe~{\sc ii} lines, some more complicated model should be applied 
to determine the AGN power law continuum, rather than to determine the 
continuum by the oversimplified wavelength windows (Kaspi et al. 2000, 
Peterson et al. 2004, Pu et al. 2006, Zhang 2013c). Due to the high noise 
of the spectra around the H$\alpha$, there are no further discussions about 
the red part of the spectra.

\begin{figure*}
\centering\includegraphics[width = 17cm,height=21cm]{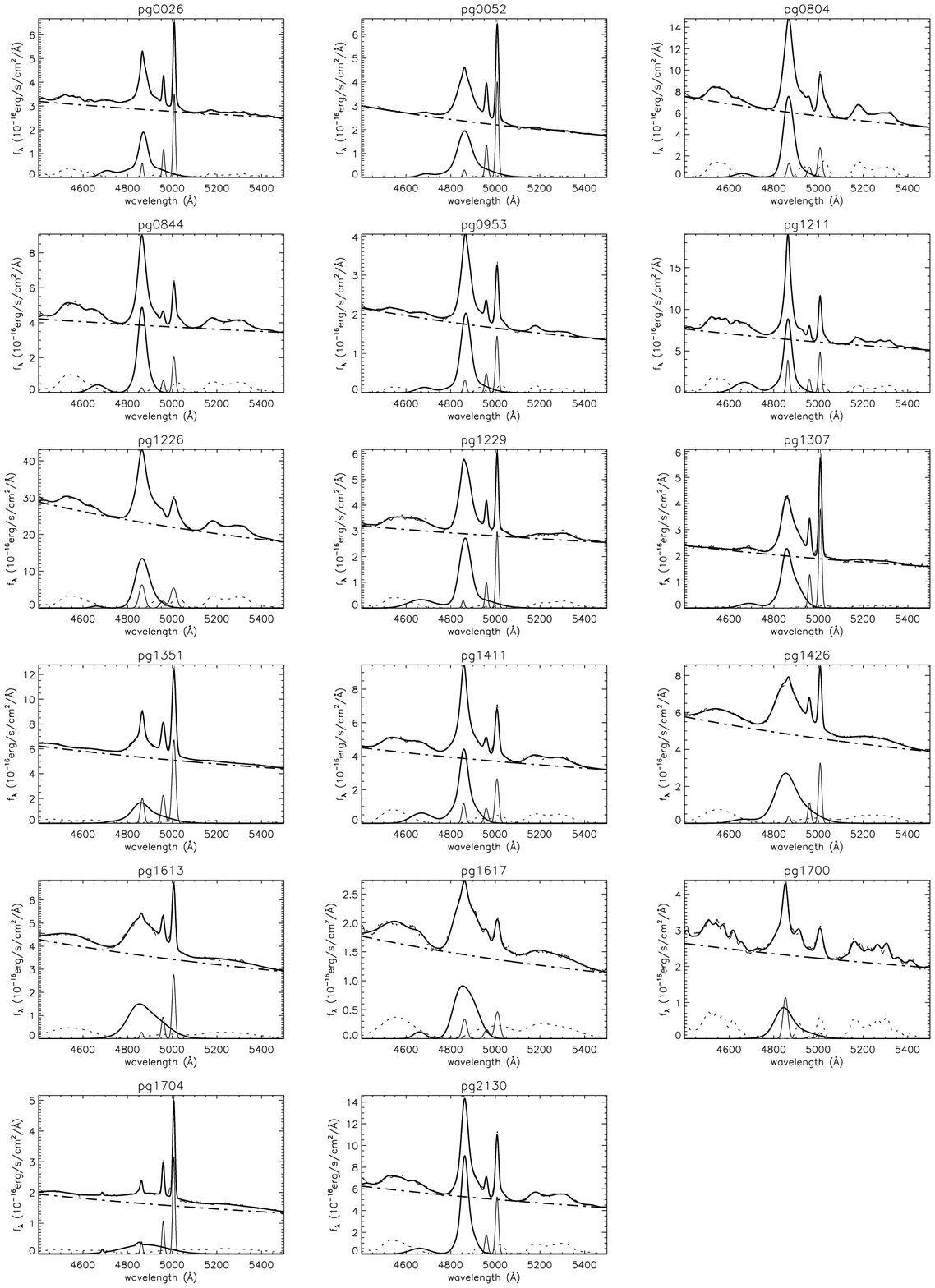}
\caption{The best fitted results for the mean spectra around the H$\beta$
of the 17 mapped PG QSOs. The thin dashed line and the thick solid line
represent the observed spectrum and the corresponding best fitted result 
respectively. Please enlarge the figure, both the thin dashed line 
and the thick solid line can be clearly detected, because the observed 
spectrum and the best fitted result are totally overlapped. Then, the 
corresponding broad H$\beta$ and probable broad He~{\sc ii} line (thick 
solid lines), narrow lines (thin solid lines), the optical Fe~{\sc ii} 
components (thin dotted line) and the power law continuum (thick 
dot-dashed line) are shown under the mean spectrum.
}
\label{fit}
\end{figure*}

   In this paper, the optical Fe~{\sc ii} lines are modeled by the more 
recent Fe~{\sc ii} template discussed in Kovacevic et al. (2010). And then, 
the power law function $f_{\lambda}\propto\lambda^{\alpha}$ is applied for 
the AGN continuum component, and multi-gaussian functions are applied to 
describe the emission lines around the H$\beta$: two broad gaussian functions 
for the broad component of the H$\beta$, one narrow gaussian function for 
the narrow H$\beta$, two narrow gaussian functions for the 
[O~{\sc iii}]$\lambda4959,5007\AA$ doublet, one broad gaussian function for 
the He~{\sc ii}$\lambda4687\AA$. In the procedure, the main objective is to 
determine the AGN power law continuum, the main reason of applying two 
broad gaussian functions for the broad H$\beta$ is only lead to the best 
fitted results for the emission lines. No further discussions are shown 
about the multiple components of the broad H$\beta$. Then, through the 
Levenberg-Marquardt least-squares minimization technique, the AGN power 
law continuum, the optical Fe~{\sc ii} lines and the other emission lines 
can be well determined. Here, the best fitted results for the mean spectra 
rather than for all the observed spectra for each object are shown in 
Fig.~\ref{fit}. It is clear that our procedure to determine the AGN power 
law continuum is efficient. Finally, based on the determined power law 
function and the reported redshift for each QSO, the continuum luminosity  
at 5100\AA and the corresponding spectral index can be well calculated in 
each epoch.

    Before the end of the section, there is one point we should note. 
Based on the discussions in Maoz et al. (1994) and Kaspi et al.(2000), the 
spectrophotometric calibration has been accomplished to do the absolute 
flux calibration for each collected spectrum by the properties of the 
simultaneously observed comparison standard star. And moreover, the 
absolute flux calibration has an moderate uncertainty of $\sim10\%$, 
and the relative flux calibration uncertainty is about $\sim3\%$. Therefore, 
we do not show further discussions about the flux calibrations, but we 
accept that the value $S/N\sim10$ for the all the collected spectra, and 
we can clearly confirm that the relative flux calibration are accurate 
enough and have few effects on our determined spectra index.

\section{Optical Spectral Index - Luminosity Relationships for the PG QSOs}

     Based on the well determined optical spectral indices and the AGN 
continuum luminosities at 5100\AA, the spectral index - luminosity 
relationship can be tested for the 17 individual mapped PG QSOs. Before 
proceeding further, there is one point we should note. The following used 
continuum luminosity is not the directed measured value from the observed 
spectrum, but the value listed in the Kaspi et al. (2000) with the necessary 
corrections haveing been done for the absolute continuum flux 
inter-calibrations at 5100\AA. Moreover, the corresponding uncertainty 
for the continuum flux in each epoch is the value listed in Kaspi et al. (2000). 

\begin{figure*}
\centering\includegraphics[width = 18cm,height=22cm]{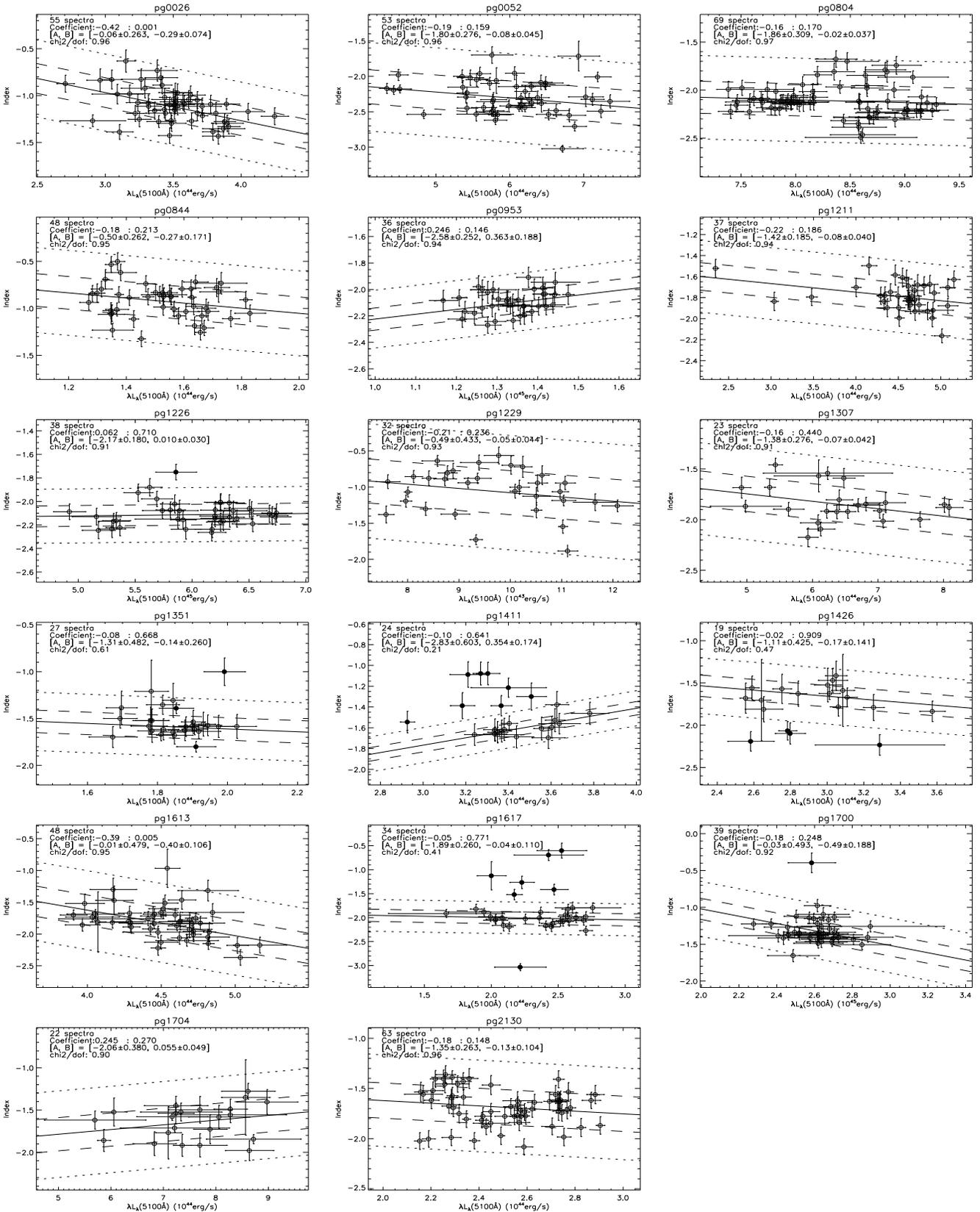}
\caption{The optical spectral index - luminosity relationships for the 
17 mapped PG QSOs. In the top left corner of each subfigure, the following 
information is shown in turn: the number of available spectra (first line), 
the corresponding spearman rank correlation coefficient and the parameter 
$P_{null}$ (second line), the parameters $A$ and $B$ 
($\alpha = A + B\times \lambda L_{\lambda}$) determined by the LTS method 
(third line), the $chi2/dof$ (forth line) for the LTS best fitted results. 
In each subfigure, the solid line represents the LTS best fitted results, 
the dashed and dotted lines are for the corresponding 68\% and 99\% 
confidence bands for the LTS best fitted results. The solid circles are 
for the outliers determined by the LTS method.
}
\label{index_lum}
\end{figure*}

\begin{figure*}
\centering\includegraphics[width = 18cm,height=22cm]{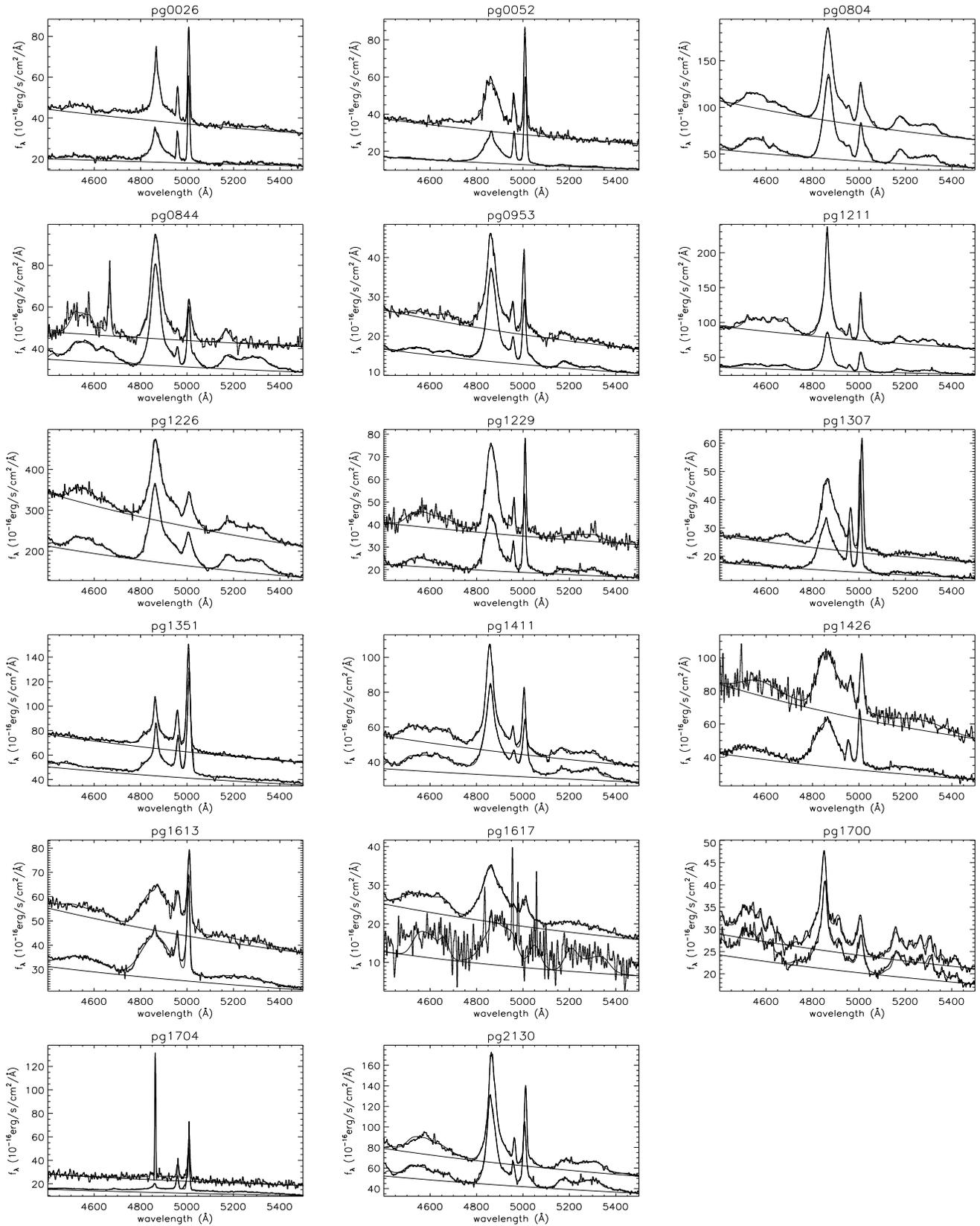}
\caption{The spectra with the lowest and the highest continuum luminosity 
at 5100\AA for each PG QSO. The thin and thick solid lines are for the 
observed spectrum and the corresponding best fitted result respectively. 
The solid line under the observed spectrum is for the determined power 
law AGN continuum.
}
\label{index_sp}
\end{figure*}

    Fig.~\ref{index_lum} shows the optical spectral index - luminosity 
relationships for the 17 individual PG QSOs. The corresponding spearman 
rank correlation coefficient and the number of the available observed 
spectra are marked for each QSO in the figure. It is clear that there are 
no apparently strong correlations between the optical spectral index and the 
continuum luminosity for the mapped QSOs, besides weak trends for the 
relationship (coefficient still larger than -0.5 but the two-sided 
significance level $p_{null}$ much smaller than 0.01) for the two PG QSOs 
of PG 0026 and PG 1613.  Besides the direct shown optical spectral index - 
luminosity relationship for the 17 mapped PG QSOs, the spectra with the 
lowest and the highest continuum luminosities at 5100$\AA$ for each QSO 
are shown in Fig.~\ref{index_sp}, in order to more clearly show spectral 
variabilities. From the results in  Fig.~\ref{index_lum} and 
Fig.~\ref{index_sp}, the basic and clear results can be found that there 
are apparent continuum variability but the much weak corresponding 
spectral index variability.

    Before the end of the section, two methods are applied to further check 
the relationships for the 17 individual QSOs: the commonly used bootstrap 
method to estimate the confidence levels of the coefficients and the more 
recent Least Trimmed Squares (LTS) robust fit method (Cappellari et al. 
2013) to determine the best fitted results for the relationships.

\begin{figure*}
\centering\includegraphics[width = 18cm,height=22cm]{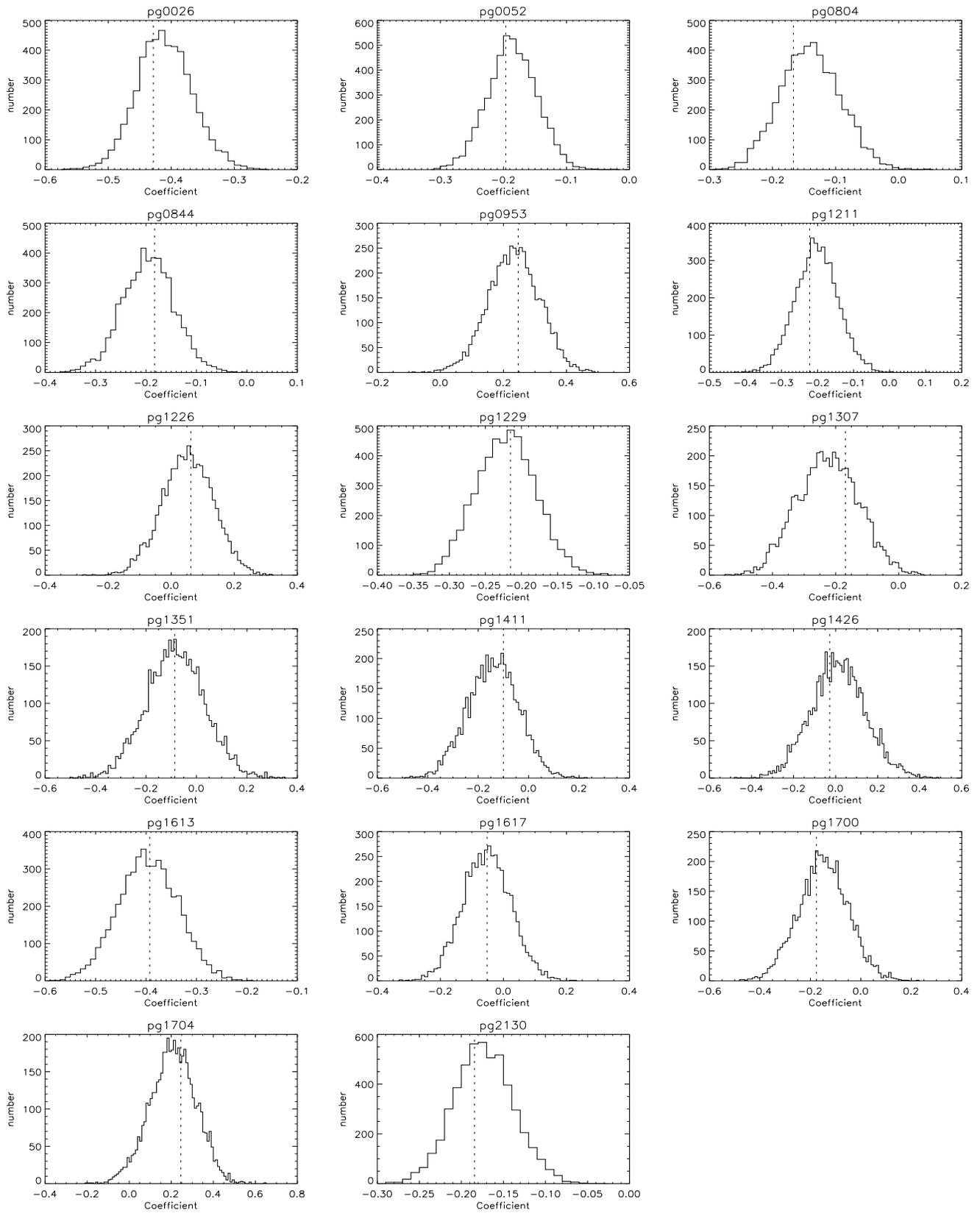}
\caption{The distributions of the spearman rank correlation coefficients 
for the relationships between the optical spectral index and the continuum 
luminosity, with the considerations of the parameter uncertainties.
}
\label{re_coe}
\end{figure*}

    The commonly used bootstrap method is applied as  follows. Before the 
spearman rank correlation coefficient is calculated, the values of the 
optical spectral index and the continuum luminosity are re-calculated and 
randomly determined within the range from $P-P_{err}$ to $P+P_{err}$, 
where $P$ and $P_{err}$ represent the parameter value (the optical spectral 
index, continuum luminosity) and the corresponding uncertainty. Then, 
the spearman rank coefficients are re-calculated by the new values of 
the optical spectral index and the continuum luminosity. The the procedure 
is repeated 5000 times for each QSO. The probability distributions of 
the re-calculated spearman rank correlation coefficients are shown in 
Fig.~\ref{re_coe} for the 17 mapped PG QSOs. It is clear that even after 
the considerations of the parameter uncertainties, the previous results 
can not be changed: no apparently strong spectral index -luminosity 
relationship for the QSOs (the coefficient always larger than -0.5 and 
smaller than 0.5). 
       
     The more recent Least Trimmed Squares (LTS) robust fit method is
applied to find the best fits for the correlations (Cappellari et al. 2013),
$\alpha = A + B\times \lambda L_{\lambda}$, under the considerations of 
the probable intrinsic scatters of the data points with uncertainties in 
both coordinates. The best fitted LTS results and the corresponding 
68\% and 99\% confidence bands for the best fitted results are shown 
in Fig.~\ref{index_lum}.  It is more clearer that besides the PG 0026 and 
PG 1613, none of the QSOs has the slope $\alpha$ 3 times larger than the 
corresponding uncertainty of the slope.  Actually, through the LTS method, 
some outliers (having deviations larger than $2.6\sigma$ from the expected 
relation) (solid circles in Fig~\ref{index_lum}) should be firstly 
ruled out, before to give the final best fitted results, which should 
lead to some more apparent linear fit. The outliers are perhaps due to 
the bad spectra quality. Meanwhile, if the outliers were also considered, 
the slope should be more closer to zero.

   Therefore, none of the 17 QSOs have apparent optical spectral index - 
luminosity relationships with $5\sigma$ confidence levels, and only two 
QSOs (PG 0026 and PG 0052) have probable and weak optical spectral index - 
luminosity relationship with $3\sigma$ confidence levels.

\section{Discussions and Conclusions}
    
     Although, there is so far no confirmed conclusion about the nature 
of the AGN variabilities, the more recent proposed damped random walk (DRW) 
model (one special stochastic model) has be well and successfully applied to 
describe the AGN variabilities (Bauer et al. 2009, Kelly et al. 2009, 
Kozlowski et al. 2010, Schmidt et al. 2010, MacLeod et al. 2010, 
Meusinger et al. 2011, Bailer-Jones 2012, MacLeod et al. 2012, 
Schmidt et al. 2012, Andrae et al. 2013, Zhang 2013b, Zu et al. 2013). 
The basic idea of the DRW model is that the variability $s(t)$ has one 
simple exponential covariance between two different epochs $t_i$ and $t_j$, 
$<s(t_i)s(t_j)>=\sigma^2\times exp(-|t_i-t_j|/\tau_0)$. Then, through the 
two DRW parameters, the damped intrinsic variability time scale $\tau_0$ 
and the damped intrinsic variability amplitude $\sigma$, the AGN 
variabilities in both observed and unobserved epochs can be well re-produced. 
And moreover, the determined damped intrinsic variability time scale 
$\tau_0$ can be used to understand the origination (or the principal 
dependent AGN 
parameter) of AGN variability, comparison with the theoretical characteristic 
time scales (Edelson \& Nandra 1999). Therefore, the variability 
properties of the optical spectral index and the continuum emission are 
checked: much different (similar) damped intrinsic variability time scales 
should lead to weak (strong) dependence of the optical spectral index on 
the continuum luminosity.

   Based on the well determined spectral index and the continuum emission 
in each epoch, the time dependent variabilities of the optical spectral 
index and the continuum emission for each PG QSO can be well analyzed by 
the well applied damped random walk method. Here, we apply the DRW 
method discussed in Zu et al. (2011, 2013) to describe the variabilities. 
In the DRW method, through the MCMC (Markov Chain Monte Carlo) analysis with the 
uniform logarithmic priors of the DRW parameters of $\tau_0$ and $\sigma$ 
covering every possible corner of the parameter space ($0<\tau_0<1e+5$ and 
$0<\sigma<1e+2$), the posterior distributions of the DRW parameters can be 
well determined and provide the final accepted parameters and the 
corresponding statistical confidence limits. Then, based on the posterior 
distributions, the exponential covariance for variability is applied to 
produce the corresponding variability at any epoch, i.e., the mean DRW fit  
and the corresponding $1\sigma$ variance. Moreover, when the DRW method is 
applied, one gaussian white noise with zero mean and unit standard deviation 
is accepted as the model measurement noise. Fig.~\ref{drw} shows the 
mean DRW fits for the variabilities of the optical spectral index and 
the continuum emission at 5100$\AA$, and Fig.~\ref{drw_par} shows the 
posterior distributions of the damped intrinsic variability time scales.  
Before further discussions about the damped intrinsic variability time 
scales, we can find that the mean DRW fits look bad for the variations of 
the spectral index in several cases (such as for the PG 0052 and PG 0844). 
We think the bad mean DRW fits are perhaps not due to the poor damped 
random walk parameters, but seriously due to the tiny variations of the 
optical spectral index and the bad time gaps. However, even the DRW 
parameters were poor, the shown observational variability modes in Fig. 5 
could prove there are different varying modes for the spectral index 
and the continuum emission for the several bad cases: larger variability 
amplitudes for the continuum emission, but smaller amplitudes for the optical 
spectral index, and much different variation trends for the optical spectral 
index and the continuum emission.

    It is clear that the varying modes are much different for the optical 
spectral index and for the continuum emission: the damped intrinsic variability 
time scales for the optical spectral index are commonly much smaller than 
that for the continuum emission (besides the results for PG 2130), which 
strongly indicates there are much different intrinsic mechanisms 
controlling the variabilities of the optical spectral index and the 
continuum emission. Thus, it can not be naturally expected for one strong 
dependence of the optical spectral index on the continuum luminosity. 
Therefore, it is clear that except the global effects of the BH masses, 
the host galaxy contamination, the narrow line contamination etc., no 
strong dependence of the optical spectral index on the continuum luminosity 
can be confirmed for the 17 mapped individual PG QSOs. In other words, 
there is no intrinsic dependence of the optical spectral index on the 
continuum luminosity.

   Before the end of the paper, there are two points we should note. On 
the one hand, we should note that Pu et al. (2006) have discussed the 
optical spectral index - luminosity relationships for the 17 PG QSOs, and 
reported that most of the 17 objects show apparent correlations between the 
spectral slope and the continuum flux (five of them have much strong 
correlations with coefficients larger than 0.5), which is against our 
found results. The different results from Pu et al. (2006) are mainly due 
to the following two main reasons. The first main reason is due to the 
process for the continuum flux calibration, because of different instruments 
used for the spectra of the PG QSOs. In Pu et al. (2006), no further process 
is considered for the flux calibration, the values direct from the observed 
spectra were used.  The second main reason is due to the effects of the 
optical Fe~{\sc ii} lines and the probable broad 
He~{\sc ii}$\lambda4687\AA$ line on the determined spectral slope 
(Bian et al. 2010, Zhang 2013c). Thus, much different results from the 
results in Pu et al. (2006) can be found. On the other hand, in the paper, 
we firstly show the direct evidence for the much different intrinsic 
mechanisms for the variabilities of the optical spectral index and the 
continuum emission by the well applied damped random walk method. 
Therefore, no strong relationship between the optical spectral index 
and the continuum luminosity can be confirmed for the individual AGN. 
The much different damped intrinsic variability time scales are enough 
to support our final results: the common sense 'AGNs get bluer when they 
get brighter' is not so common. 
  
\vspace{8mm}
The final conclusions are as follows.
\begin{itemize}
\item In order to ignore the global effects of the host stellar lights, 
the emission lines, the central region parameters on the relationship 
between the optical spectral index and the continuum luminosity, we 
carefully analysis the multi-epoch spectra of the well known 17 individual 
mapped PG QSOs in Kaspi et al. (2000), and found there is no confirmed 
strong dependence of the optical spectral index on the continuum luminosity, 
even after the considerations of the parameter uncertainties.
\item The well accepted damped random walk method is applied for the 
variabilities of the optical spectral index and the continuum emission for 
the 17 PG QSOs, the damped intrinsic variability time scales are much 
different for the optical spectral index and the continuum emission, 
which strongly indicates there are much different mechanisms controlling 
the optical spectral index variability and the continuum emission variability.
\end{itemize}
 
\section*{Acknowledgements}
Z-XG very gratefully acknowledge Bailer-Jones C. A. L. in MPIA for 
giving us constructive comments and suggestions to greatly improve our 
paper. Z-XG gratefully acknowledges the kind support from the Chinese 
grant NSFC-11003043 and NSFC-11178003. We further gratefully thanks Dr. 
Kaspi S. who makes us to conveniently collect the public spectra of the 
17 mapped PG QSOs (http://wise-obs.tau.ac.il/\~{}shai/PG/).

\begin{figure*}
\centering\includegraphics[width = 18cm,height=22cm]{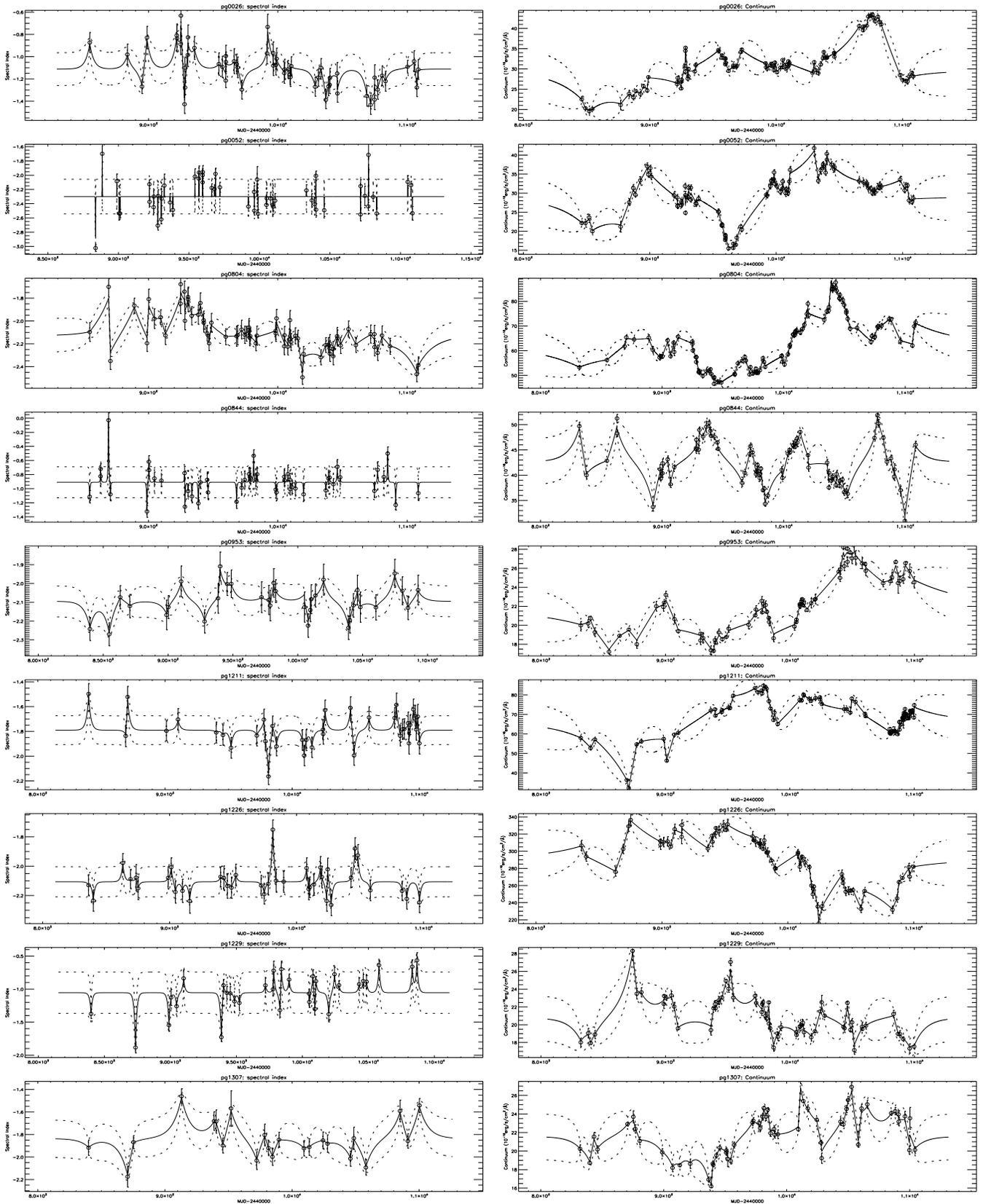}
\caption{The mean DRW fits for the variabilities of the optical 
spectral index and the continuum emission for the 17 mapped PG QSOs by 
the DRW method (Zu et al. 2011, 2013) with the DRW parameters determined 
by the MCMC analysis. The circles are for the observational data points, 
the solid line and the two dotted lines are for the determined mean DRW 
fit and the corresponding $1\sigma$ variance respectively. The left panels 
shows the results about the optical spectral index, and the right panels 
are for the continuum emission. 
}
\label{drw}
\end{figure*}

\setcounter{figure}{4}
\begin{figure*}
\centering\includegraphics[width = 18cm,height=22cm]{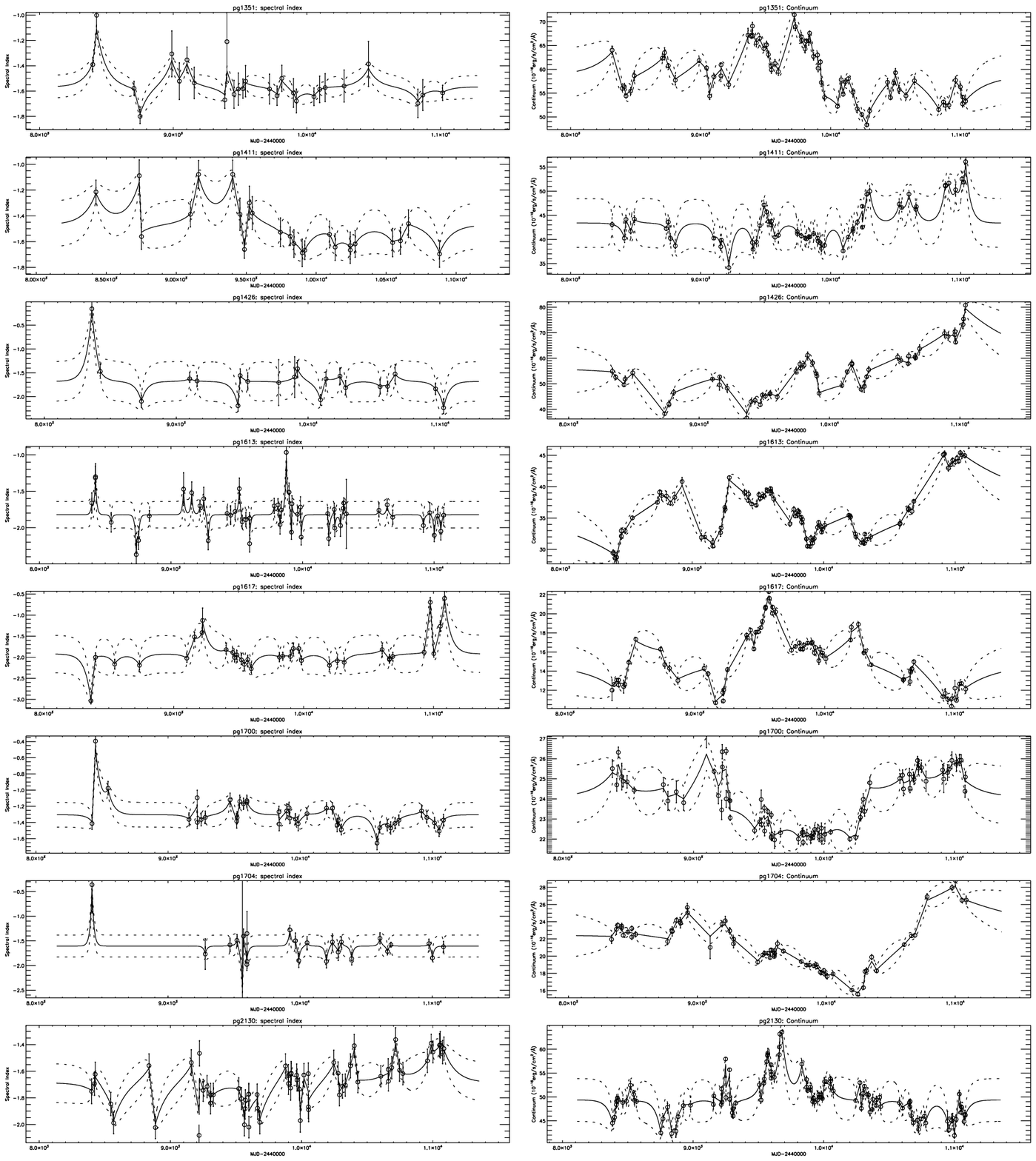}
\caption{--Continued.
}
\end{figure*}

\begin{figure*}
\centering\includegraphics[width = 18cm,height=22cm]{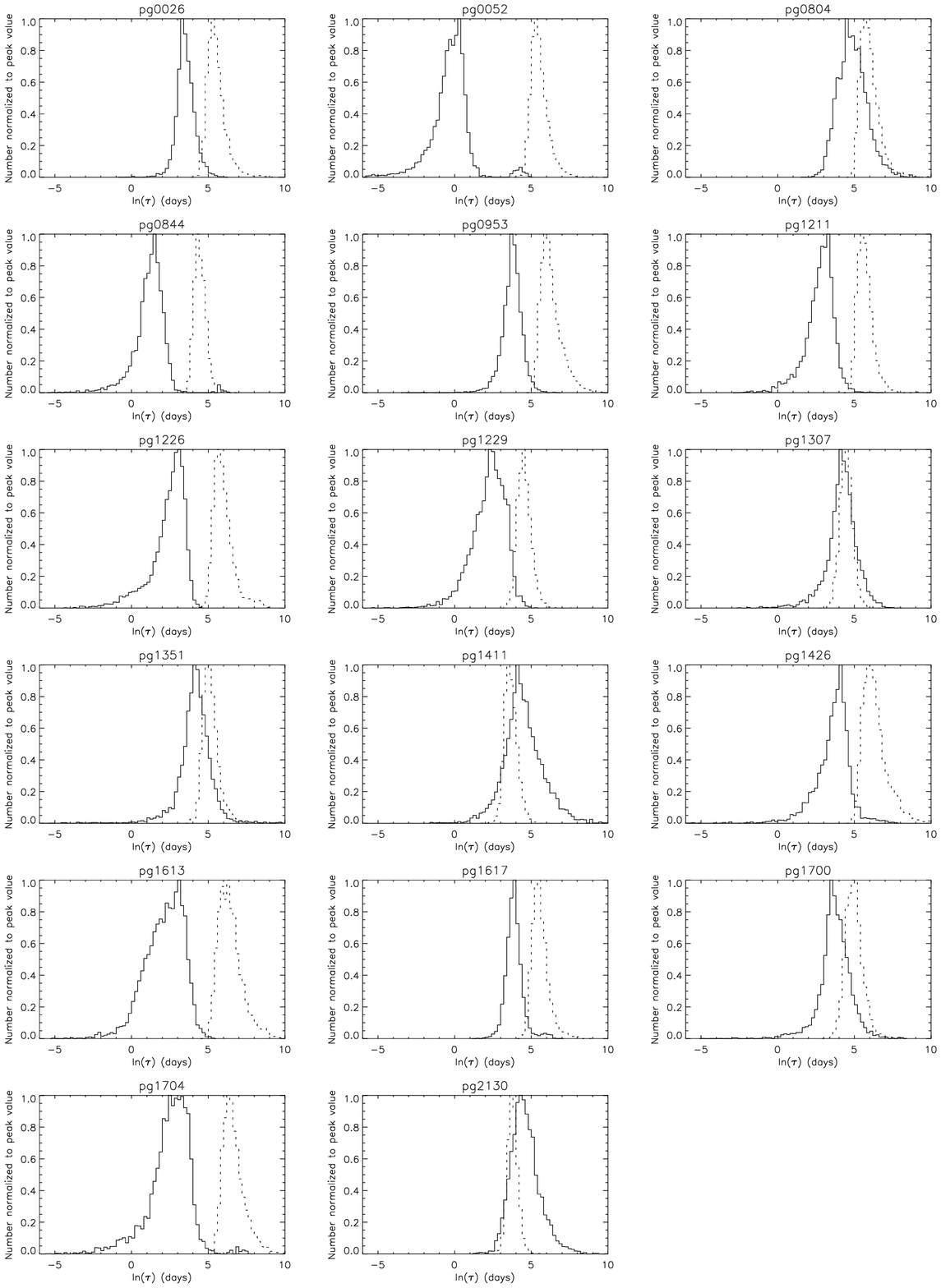}
\caption{The posterior distributions of the DRW determined damped intrinsic 
variability time scales for the variabilities of the optical spectral 
index and the continuum emission, by the MCMC analysis. Solid line 
represents the distribution for the optical spectral index variabilities, 
and the dotted line is for the continuum variability.
}
\label{drw_par}
\end{figure*}


\end{document}